\documentstyle[preprint,aps]{revtex}
\begin{document}
\draft
\title{Nuclear Dependence in Direct Photon Production}
\author{Xiaofeng Guo and Jianwei Qiu}
\address{Department of Physics and Astronomy,
         Iowa State University \\
         Ames, Iowa 50011, USA }
\date{December 7, 1995}
\maketitle

\begin{abstract}
We calculate the nuclear dependence of direct photon production in
hadron-nucleus collisions.  In terms of a multiple scattering picture,
we factorize the cross section for direct photon production into
calculable short-distance partonic parts times multiparton correlation
functions in nuclei.  We present the hadron-nucleus cross section as
$A^{\alpha}$ times the hadron-nucleon cross section.
Using information on the multiparton correlation functions
extracted from photon-nucleus experiments, we compute the value of
$\alpha$ as a function of transverse momentum of the direct photon.
We also compare our results with recent data from Fermilab experiment
E706.
\end{abstract}
\vspace{0.2in}


\section{Introduction}
\label{sec:1}

As early as in 1970s, it was observed \cite{Cronin} that inclusive
cross sections for single high transverse momentum particle
produced in hadron-nucleus scattering show an ``anomalous'' nuclear
dependence, in which the cross section at fixed transverse momentum
grows approximately as $A^{\alpha}$ with $A$ as the atomic number of
the nuclear target.  The value of $\alpha$ is a function of transverse
momentum, and can be as large as $4/3$.
This phenomenon has been known as Cronin effect.  Typical energy
exchange in a high transverse-momentum scattering process is so large
that any single hard scattering should be very localized
within a single nucleon.  Consequently, a linear $A$ dependence
is expected for single scattering processes.  Therefore, the Cronin
effect is often described as due to multiple scattering of partons in
nuclear matter \cite{Fishbane,Petersson,Bodwin,LQS}.  The $A^{4/3}$
behavior signals a dependence on nuclear size, and multiple
scattering is dominated by double scattering.

In some of the previous work on this topic, an independent scattering
picture was adopted \cite{Fishbane}.  In this picture, each scattering
was treated independently.  For example, the cross section for double
scattering was proportional to a product of two Born cross sections.
The double scattering cross section in this picture is not infrared
safe however.  This is because the kinematics of single particle
inclusive cross section can only provide a constraint on the total
momentum from the target, which leaves the
possibility that one of the Born cross section diverges when
the momentum transfer of this Born cross section approaches to
zero.  Therefore, theoretical predictions from the independent
scattering picture are sensitive to the infrared cutoffs which must be
introduced in the calculations.

Recently, Luo, Qiu and Sterman (LQS) have shown that the anomalous
nuclear enhancement can be described naturally in perturbative QCD, in
terms of a nonleading power, or ``higher twist'' formalism \cite{LQS}.
In this treatment, contribution from double scattering can be
factorized into short-distance hard parts convoluted with
corresponding multi-parton matrix elements, or
multiparton correlation functions in nuclei.  The short-distance
partonic parts are calculable in perturbative QCD, and all infrared
divergences associated with soft rescatterings in perturbation theory
can systematically be absorbed into multiparton correlation functions.
The multiparton correlation functions are nonperturbative, just like
the parton distributions in the single scattering processes.  These
correlation functions in nuclei provide information about nuclear
matter and its interaction with high energy probes.  They can reveal
information different from what normal parton distributions in
nuclei can provide; and in principle, they are as fundamental as
the parton distributions.  In order to test the theory, we need to
find different processes which depend on the same multiparton
correlation functions.  Information on these new correlation functions
extracted from one set of processes may be applied in other
processes.

Our aim in this paper is to show that the consistent perturbative QCD
treatment of double scattering developed by LQS can be naturally
applied to high transverse momentum direct photon production in
hadron-nucleus scattering.  We factorize the cross section of direct
photon production into calculable short-distance partonic parts times
multiparton correlation functions, which are the same as those derived
in Ref.~\cite{LQS}.  We calculated the short-distance partonic hard
parts.  We evaluate the nuclear dependence by using the information on
the multiparton correlation functions, extracted from
experiments on momentum imbalance of two-jet photoproduction on
nuclear targets \cite{LQS}.  Our numerical results are consistent with
recent measurement of nuclear dependence in direct photon production
from Fermilab E706 experiment \cite{Marek}.

A double scattering with high momentum transfer must have at least one
hard scattering to produce the high transverse momentum observables.
In addition, there may be a soft scattering either before or after the
hard scattering (referred to below as a soft-hard process),
or another hard scattering (called as a double hard process).
We shall show that only the soft-hard processes contribute
to the nuclear dependence of direct photon production to the order
we consider.  The fact that the photon does not interact strongly
once produced at the hard collisions eliminates final-state
multiple scattering in direct photon cross sections.  Therefore,
direct photon production in hadron-nucleus scattering provides an
excellent test of initial state multiple scatterings, while
jet or single particle production in photon-nucleus scattering
provide independent tests of final state multiple scatterings.
Jet and single particle production in hadron-nucleus collisions, on
the other hand, receive contributions from both initial and
final state multiple scatterings.  Final state multiple scattering in
photoproduction has been discussed in Ref.~\cite{LQS}.  Our work will
provide the complimentary information on the initial state multiple
scattering.

We begin in Sec.~\ref{sec:2} with an outline of the formalism used in
our calculation.  Complete analytical results of our calculation are
also presented in Sec.~\ref{sec:2}.  The detailed derivation of our
results, and the calculation of the partonic hard parts are presented
in Sec.~\ref{sec:3}.  In Sec.~\ref{sec:4}, we present our numerical
results.  We also compare our numerical results with recent
experimental data.  We conclude with a brief summary, and suggestions
for further work.


\section{Formalism and Analytical Results}
\label{sec:2}

An energetic photon can be directly produced at short-distance in high
energy collisions, and does not interact strongly once produced.
Therefore, it has been recognized for a long time that direct photon
production is a clean probe for short-distance dynamics in high
energy collisions \cite{Owens}.  Data from hadronic prompt photon
production plays a very important role in QCD global analysis, and
provides constraints on gluon distributions in hadrons
\cite{ABFOW,CTEQ3,MRS}.
In this section, we outline the general formulas for cross sections of
direct photon production, and present our analytical results for the
contribution from double scattering processes.

\subsection{Formalism}
\label{sec:2a}

In the following discussion, we study direct photon production in
hadron-nucleus collisions,
\begin{equation}
h(p')+A(p) \longrightarrow \gamma (l) + X\ ,
\label{e1}
\end{equation}
where $p$ is defined as the averaged momentum per nucleon.  In
general, the total cross section for the above process can be
expressed as a sum of contributions from single scattering, double
scattering and even higher multiple scattering,
\begin{equation}
d\sigma_{hA\rightarrow\gamma}(l)
 = d\sigma_{hA\rightarrow\gamma}^{(S)}(l)
 + d\sigma_{hA\rightarrow\gamma}^{(D)}(l)+ ... \ ,
\label{e2}
\end{equation}
where the superscripts $(S)$ and $(D)$ represent the single and double
scattering, respectively, and ``...'' represents other possible
multiple scatterings.  In this paper, we consider only the double
scattering, and its contribution to nuclear dependence.

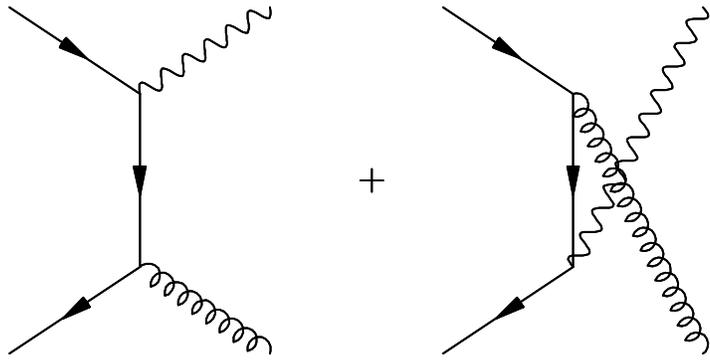
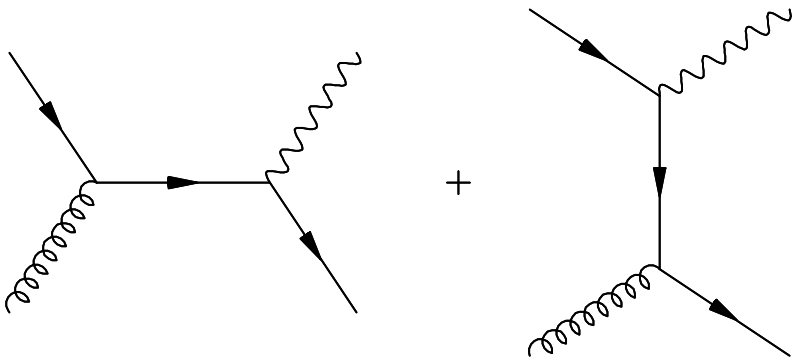
\begin{figure}
\caption{Lowest order Feynman diagrams contribute to single
scattering: a) ``Annihilation'', b) ``Compton''.}
\label{fig1}
\end{figure}

As a result of perturbative factorization \cite{CSS}, the single
scattering cross section can be expressed as
\begin{eqnarray}
d\sigma_{hA\rightarrow\gamma}^{(S)}(l)
&=& A d\sigma_{hN\rightarrow\gamma}^{(S)}(l) \nonumber \\
&=& A\sum_{a,b}\int dx'\, f_{a/h}(x')\, \int dx\, f_{b/N}(x)\,
     d\hat{\sigma}_{ab\rightarrow\gamma}(x',x,l)\ .
\label{e3}
\end{eqnarray}
In Eq.~(\ref{e3}), $f_{a/h}(x')$ is a normal parton distributions in
the beam hadron $h$, $f_{b/N}(x)$ is an effective nucleon parton
distributions inside a nucleus, which should include the well-known
EMC effect.  In principle, the parton-parton scattering cross section,
$d\hat{\sigma}_{ab\rightarrow\gamma}$, should include both direct and
fragmentation contributions.  That is, an energetic photon can be
produced directly at short-distances, or produced from the
fragmentation of an energetic parton which was produced at
short-distances \cite{Owens,BQ}.  For example, the partonic scattering
may produce an energetic quark, which radiates a photon.
Since we are most interested in fixed target
experiments here, the fragmentation contribution is much smaller than
the direct contribution in most of phase space \cite{BQ}.
Therefore, in the rest of our discussion, we will consider only the
direct production of photons.  For example, at the lowest order, we
have contributions from
$q\bar{q}\rightarrow\gamma g$ ``Annihilation'' diagrams,
sketched in Fig.~\ref{fig1}a; and $gq(\mbox{or\ }\bar{q})
\rightarrow\gamma q(\mbox{or\ }\bar{q})$ ``Compton'' diagrams,
sketched in Fig.~\ref{fig1}b.

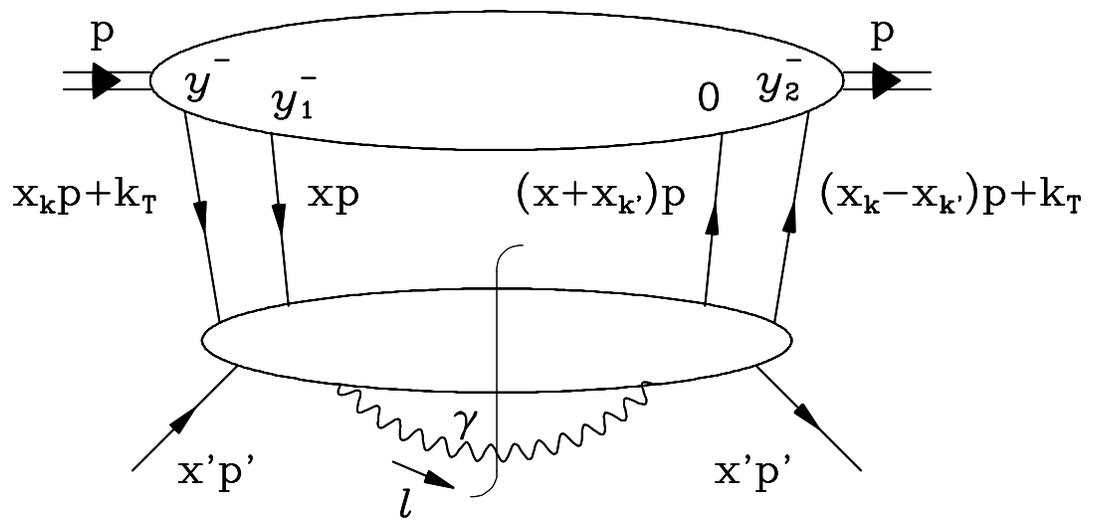
\begin{figure}
\caption{A Graphical representation of double scattering contributions
from the parton-nucleus collisions.}
\label{fig2}
\end{figure}

In terms of the generalized factorization theorem \cite{QS},
the double scattering cross section can be written as:
\begin{equation}
d\sigma^{(D)}_{hA\rightarrow\gamma}(l)
= \sum_{a} \int dx' f_{a/h}(x')\,
  d\sigma^{(D)}_{aA\rightarrow\gamma}(x',p,l) \ ,
\label{e4}
\end{equation}
where $d\sigma^{(D)}_{aA\rightarrow\gamma}(x',p,l)$ can be thought as
the double scattering cross section between a parton and the nucleus.
At the lowest order, it can be factorized as
\begin{equation}
d\sigma^{(D)}_{aA\rightarrow\gamma}(x',p,l)
= \int dx\, dx_{k}\, dx_{k'} \sum_{\{i\}}
  T_{\{i\}}(x,x_{k},x_{k'})\, H_{\{i\}}(x',x,x_{k},x_{k'},l)\ .
\label{e5}
\end{equation}
In Eq.~(\ref{e5}), $T_{\{i\}}(x,x_{k},x_{k'})$ are the matrix elements
of four-parton operators, characterized by the set of fields operators
$\{i\}$; and $H_{\{i\}}$ are the corresponding partonic hard
scattering functions.  The $x$,
$x_k$ and $x_{k'}$ are independent collinear momentum fractions
carried by the partons from the nucleus.  The graphical representation
of Eq.~(\ref{e5}) is shown in Fig.~{\ref{fig2}}.  At the lowest order,
there are three types of partonic subprocesses that contribute to the
double scatterings.  Feynman diagrams of these partonic subprocesses
are sketched in Fig.~\ref{fig3}.

We define the invariant direct photon cross section in hadron-nucleus
collision in terms of cross sections in hadron-nucleon collisions,
\begin{eqnarray}
E_{l}\frac{d\sigma_{hA\rightarrow\gamma}(l)}{d^{3}l}
&\equiv & A^{\alpha(l)}\,
          E_{l}\frac{d\sigma_{hN\rightarrow\gamma}^{(S)}(l)}
                    {d^{3}l} \nonumber \\
&\approx & E_{l}\frac{d\sigma_{hA\rightarrow\gamma}^{(S)}(l)}
                     {d^{3}l}
         + E_{l}\frac{d\sigma_{hA\rightarrow\gamma}^{(D)}(l)}
                     {d^{3}l}\ ,
\label{e6}
\end{eqnarray}
where Eq.~(\ref{e2}) was used.
Substituting Eq.~(\ref{e3}) into Eq.~(\ref{e6}), we obtain the
definition for the nuclear dependence parameter $\alpha(l)$,
\begin{equation}
\alpha(l)=1+\frac{1}{\ell n(A)}\,
   \ell n\left(1+\frac{1}{A}\,
               \frac{E_{l}\frac{d\sigma^{(D)}_{hA\rightarrow\gamma}(l)}
                               {d^{3}l}}
                    {E_{l}\frac{d\sigma^{(S)}_{hN\rightarrow\gamma}(l)}
                               {d^{3}l}}
         \right)\ .
\label{e7}
\end{equation}
{}From Eq.~(\ref{e7}), $\alpha(l)>1$ if
$d\sigma^{(D)}_{hA\rightarrow\gamma}/d^3l$ is positive, which will
turn out to be the case for the kinematic regime
in which we are interested here.  However, in
general, the double scattering contribution $\sigma^{(D)}$ may be
negative, and $\alpha(l)<1$ in certain part of phase space.  The
positivity of a cross section requires the sum of all possible
multiple scattering contribution to be positive.  The separation
between single and double scattering is not unique.  For example,
two scatterings can be very close to each other, and localized in one
nucleon, and such a double scattering will not provide the anomalous
nuclear dependence, and may be classified as a single scattering.

We will argue later that the leading double scattering contribution,
$d\sigma^{(D)}_{hA\rightarrow\gamma}/d^3l$, is proportional to
$A^{4/3}$.  Consequently, the value of $\alpha(l)$ will be between $1$
and $4/3$, depending on the relative sizes of contributions from the
single and double scatterings.  If the double scattering contribution
is larger than the single scattering contribution in a certain
part of the phase space, the value of $\alpha(l)$ in that region will
approach $4/3$.

\subsection{Analytic Results}
\label{sec:2b}

In this subsection, we present the analytic results which are used
to calculate the nuclear dependence parameter $\alpha(l)$ defined in
Eq.~(\ref{e7}).

Following Eq.~(\ref{e3}), the lowest order invariant cross section for
single scattering direct photon production is given by \cite{Owens}
\begin{eqnarray}
E_{l}\frac{d\sigma^{(S)}_{hN\rightarrow\gamma}(l)}{d^{3}l}
&=& \sum_{a,b}\int dx'\, f_{a/h}(x')\,
              \int dx\, f_{b/N}(x)\,
              \delta\left(x-\frac{-x't}{x's+u}\right) \nonumber \\
&\times & \alpha_{em}\, \alpha_s\, \left(\frac{1}{\hat{s}}\right)
              \left(\frac{1}{x's+u}\right)\,
              \left|\overline{M}_{ab\rightarrow\gamma}\right|^2 \ ,
\label{e8}
\end{eqnarray}
where $\sum_{a,b}$ run over all gluon, quark and antiquark flavors;
and the matrix elements for the ``Annihilation'' and ``Compton''
subprocesses, sketched in Fig.~\ref{fig1}, are given by
\begin{mathletters}
\label{e9}
\begin{eqnarray}
\left|\overline{M}_{q\bar{q}\rightarrow\gamma g}\right|^2
&=&e_q^2\, \left(\frac{4}{9}\right)\,
        2\ \left(\frac{\hat{u}}{\hat{t}} +
                 \frac{\hat{t}}{\hat{u}}\right)\ ;
\label{e9a} \\
\left|\overline{M}_{qg\rightarrow\gamma q}\right|^2
&=&e_q^2\, \left(\frac{1}{6}\right)\,
        2\ \left(\frac{-\hat{t}}{\hat{s}} +
                 \frac{\hat{s}}{-\hat{t}}\right)\ ;
\label{e9b}
\end{eqnarray}
\end{mathletters}

\noindent where $e_q$ is the fractional charge carried by a quark of
type ``$q$''.
The invariants $\hat{s}$, $\hat{t}$ and $\hat{u}$ are usual Mandelstam
invariants for the parton-parton subprocess.  They are related to
those at the hadron-nucleon interaction by
\begin{mathletters}
\label{e10}
\begin{eqnarray}
\hat{s}=x'\, x\, s, &\hspace{0.3in} \hat{t}=x'\, t,
                    &\hspace{0.3in} \hat{u}=x\, u\ ;
\label{e10a} \\
s=(p'+p)^2,         &\hspace{0.3in} t=(p'-l)^2,
                    &\hspace{0.3in} u=(p-l)^2 \ .
\label{e10b}
\end{eqnarray}
\end{mathletters}

In the case of double scattering, there are four physical partons
linking the matrix elements $T$ and the partonic hard parts $H$, as
shown in Eq.~(\ref{e5}).  After taking into account of momentum
conservation, there are still \underline{three} independent momentum
fraction integrations ($x$, $x_k$ and $x_{k'}$ defined in
Fig.~\ref{fig2}) between the matrix elements and the partonic parts,
in contrast to \underline{one} independent momentum fraction
integration for the case of single scattering.  As explained in next
section, we take the leading pole approximation to integrate over two
of the three momentum fractions (e.g., $x_k$ and $x_{k'}$).  Then, the
invariant double scattering cross section
$E_{l}d\sigma^{(D)}_{hA\rightarrow\gamma}/d^{3}l$ can be reduced into
a form very similar to the single scattering cross section defined in
Eq.~(\ref{e8}).  Following our derivation in next section, we
obtain
\begin{eqnarray}
E_{l}\frac{d\sigma^{(D)}_{hA\rightarrow\gamma}(l)}{d^{3}l}
&=& \alpha_{em}(4\pi\alpha_{s})^{2}
\int dx'\, dx\, \delta\left(x-\frac{-x't}{x's+u}\right)
\left(\frac{1}{x's}\right) \left(\frac{1}{x's+u}\right) \nonumber \\
&\times & \sum_{q} e_{q}^{2}
\left[\ f_{\bar{q}/h}(x')\, \Phi_{q}(x,x',A)\, H_{q\bar{q}} \right.
      \nonumber \\
&& {\hskip 0.32in}
      + f_{q/h}(x')\, \Phi_{g}(x,x',A)\, H_{g} \nonumber \\
&& {\hskip 0.32in}
\left. + f_{g/h}(x')\, \Phi_{q}(x,x',A)\, H_{q}\ \right] \ ,
\label{e11}
\end{eqnarray}
where $\sum_{q}$ runs over all quark and antiquark flavors.
In Eq.~(\ref{e11}), the functions $\Phi_{i}$ with $i=q,g$ represent
the effective parton flux from the nucleus.  They are given by
\begin{equation}
\Phi_{i} = \left[ \frac{\partial^{2}}{\partial x^{2}}
\left( \frac{T_{i}(x,A)}{x} \right) \right]
\left( \frac{l_{T}^{2}}{(x's+u)^{2}} \right)
+\left[ \frac{\partial}{\partial x}
\left( \frac{T_{i}(x,A)}{x} \right) \right]
\left( \frac{-u}{x's(x's+u)} \right) \ .
\label{e12}
\end{equation}
The $T_{i}(x,A)$ with $i=q,g$ in Eq.~(\ref{e12}) are the twist-four
matrix elements in nuclei. They were originally introduced in
Ref.~\cite{LQS}, and are given by
\begin{mathletters}
\label{e13}
\begin{eqnarray}
T_{q}(x,A) &=&
 \int \frac{dy_{1}^{-}}{2\pi}e^{ixp^{+}y_{1}^{-}}
 \int \frac{dy^{-}dy_{2}^{-}}{2\pi}
      \theta(y_{1}^{-}-y^{-})\theta(-y_{2}^{-}) \nonumber \\
&\ & \times \frac{1}{2}
     \langle p_{A}|F_{\alpha}^{\ +}(y_{2}^{-})\bar{\psi}_{q}(0)
                  \gamma^{+}\psi_{q}(y_{1}^{-})F^{+\alpha}(y^{-})
     |p_{A}\rangle\ ;
\label{e13a}
\end{eqnarray}
and
\begin{eqnarray}
T_{g}(x,A) &=&
 \int \frac{dy_{1}^{-}}{2\pi}e^{ixp^{+}y_{1}^{-}}
 \int \frac{dy^{-}dy_{2}^{-}}{2\pi}
      \theta(y_{1}^{-}-y_{2}^{-})\theta(-y^{-}) \nonumber \\
&\ & \times \frac{1}{xp^{+}}
     \langle p_{A}|F^{\sigma+}(y_{2}^{-}) F_{\alpha}^{\ +}(0)
                   F^{+\alpha}(y_{1}^{-})F_{\ \sigma}^{+}(y^{-})
     |p_{A}\rangle\ .
\label{e13b}
\end{eqnarray}
\end{mathletters}
\noindent In Eq.~(\ref{e13}), $F_{\mu\nu}$ and $\psi_q$ are the field
strength and quark field operator, respectively.

In Eq.~(\ref{e11}), the $H_{i}$ are the partonic hard parts, and
\begin{mathletters}
\label{e14}
\begin{eqnarray}
H_{q\bar{q}}&=&\left( \frac{2}{27} \right)
        \left( \frac{-u}{x's+u}+\frac{x's+u}{-u} \right)\ ;
\label{e14a} \\
H_{g}&=&\left( \frac{1}{36} \right)
        \left( \frac{x's}{x's+u}+\frac{x's+u}{x's} \right)\ ;
\label{e14b} \\
H_{q}&=&\left( \frac{1}{16} \right)
        \left( \frac{x's}{-u}+\frac{-u}{x's} \right)\ ,
\label{e14c}
\end{eqnarray}
\end{mathletters}
which we will derive in next section.

Eqs.~(\ref{e11}), (\ref{e12}) and (\ref{e14}) are our complete
analytic results at leading nonvanishing order in $\alpha_s$.
As usual, the next-to-leading order (NLO)
contribution might be important for the single and/or double
scatterings.  However, since the nuclear dependence
parameter $\alpha(l)$, defined in Eq.~(\ref{e7}), depends on the ratio
of the double and single scattering contributions, we expect that the
values of $\alpha(l)$ presented in this paper are not very sensitive to
the NLO contributions.


\section{Derivation of the Double Scattering Contributions}
\label{sec:3}

In this section we provide the derivation that leads to the analytic
results presented in last section.  The method that we used here was
first introduced in Ref.~\cite{LQS}.  It can be summarized in
following technical steps: a) factorize the double scattering
contribution into a convolution between the partonic hard parts and the
corresponding multiparton matrix elements (e.g., see Eq.~(\ref{e5}));
b) in the leading pole approximation, integrate over two of the three
independent momentum fractions by contour integrations, and reexpress
the multiparton matrix elements in terms of the $T_q(x,A)$ and
$T_g(x,A)$ defined in Eq.~(\ref{e13}); c) calculate the corresponding
partonic hard parts.

At lowest order, only \underline{three} types of partonic
subprocesses, as sketched in Fig.~\ref{fig3}, contribute to the double
scattering cross section $d\sigma^{(D)}_{aA\rightarrow\gamma}$
introduced in Eq.~(\ref{e5}).  These three subprocesses correspond to
adding two gluons to the lowest order ``Annihilation'' and ``Compton''
subprocesses, shown in Fig.~\ref{fig1}.  In the following subsections,
we present the detailed derivation for one subprocess, and provide the
results for other subprocesses.

\subsection{Perturbative factorization}
\label{subsec:3a}

Consider the subprocess shown in Fig.~\ref{fig3}a, in which there are
\underline{three} independent \underline{four-momentum} linking the
partonic part and corresponding two-quark-two-gluon matrix element.
In the center of mass frame of high energy collision, all partons
inside the nucleus are moving almost parallel to each other, along the
direction of the nucleus.  Therefore, all \underline{three} parton
momenta can be approximately replaced by the components collinear to
the hadron momentum.  After such a collinear expansion, the double
scattering contribution from the generalized ``Annihilation''
subprocess shown in Fig.~\ref{fig3}a can be written as \cite{LQS}
\begin{equation}
E_{l}\frac{d\sigma^{(D)}_{qA\rightarrow\gamma}}{d^{3}l}
= \frac{1}{2x's}\,\int dx\, dx_{k}\, dx_{k'}\,
\int d^{2}k_{T}\, \overline{T}(x,x_{k},x_{k'},k_{T},p)\,
\bar{H}(x'p',x,x_{k},x_{k'},k_{T},p,l)\ ,
\label{e15}
\end{equation}
where $2x's$ is the flux factor between the incoming beam quark
and the nucleus, and $x'p'$ is the momentum carried by the beam quark.
In Eq.~(\ref{e15}), the two-quark-two-gluon matrix
element, $\overline{T}$, is defined as
\begin{eqnarray}
\overline{T}(x,x_{k},x_{k'},k_{T},p)
&=& \int \frac{dy_{1}^{-}}{2\pi} \frac{dy_{2}^{-}}{2\pi}
    \frac{dy^{-}}{2\pi} \frac{d^{2}y_{T}}{(2\pi)^2} \nonumber \\
&\ & \times e^{ixp^{+}y_{1}^{-}}\, e^{ix_{k}p^{+}y^{-}}\,
            e^{-i(x_{k}-x_{k'})p^{+}y_{2}^{-}}\,
            e^{-ik_{T}\cdot y_{T}} \nonumber \\
&\ & \times \frac{1}{2}\langle p_{A} |
            A^+(y_{2}^{-},0_{T})\, \bar{\psi}_q(0)\, \gamma^+ \,
            \psi_q(y_{1}^{-})\, A^+(y^{-},y_{T}) | p_{A}\rangle\ .
\label{e16}
\end{eqnarray}
The corresponding partonic part $\bar{H}$ is given by the
diagrams shown in Fig.~\ref{fig4}, with gluon lines contracted with
$p^{\rho}p^{\sigma}$, quark lines from the target traced with
$(\gamma\cdot p)/2$, and quark lines from the beam traced with
$(\gamma\cdot (x'p'))/2$.  Here, we work in Feynman gauge,
in which the leading contribution from the gluon field operators is
$A^{\rho}\approx A^+(p^{\rho}/p^+)$.  We also kept the $k_T$ for the
gluons in order to extract a double scattering contribution beyond
leading twist.

By expanding the partonic part $\bar{H}$ introduced in
Eq.~(\ref{e15}) at $k_T=0$, we have
\begin{eqnarray}
\bar{H}(x'p',x,x_{k},x_{k'},k_{T},p,l)
&=& \bar{H}(x'p',x,x_{k},x_{k'},k_{T}=0,p,l) \nonumber \\
&+& \left. \frac{\partial \bar{H}}{\partial k_{T}^{\alpha}}
    \right|_{k_{T}=0}\ k_{T}^{\alpha}\,
+\, \left. \frac{1}{2}\, \frac{\partial^{2}\bar{H}}
                {\partial k_{T}^{\alpha} \partial  k_{T}^{\beta}}
    \right|_{k_{T}=0}\ k_{T}^{\alpha}\, k_{T}^{\beta} + \ldots\ .
\label{e17}
\end{eqnarray}
In the right-hand-side of Eq.~(\ref{e17}), the first term is the
leading twist eikonal contribution, which does not correspond to
physical double scattering, but simply makes the single-scattering
matrix element gauge invariant.
The second term vanishes after integrating over $k_T$.  The
third term will give the finite contribution to the multiple
scattering process.  Substituting Eq.~(\ref{e17}) into
Eq.~(\ref{e15}), and integrating over $d^2k_T$, we obtain
\begin{eqnarray}
E_{l}\frac{d\sigma^{(D)}_{qA\rightarrow\gamma}}{d^{3}l}
&=& \frac{1}{2x's}\,
\int dx\, dx_{k}\, dx_{k'}\, T(x,x_k,x_{k'},A) \nonumber \\
&\times &
\left(-\frac{1}{2}g^{\alpha\beta}\right)
\left[\frac{1}{2}\, \frac{\partial^{2}}
                   {\partial k_{T}^{\alpha} \partial k_{T}^{\beta}}
    \bar{H}(x'p',x,x_k,x_{k'},k_T=0,p,l) \right]\ ,
\label{e18}
\end{eqnarray}
where the modified matrix element $T$ is given by
\begin{eqnarray}
T(x,x_k,x_{k'},A) &=&
\int \frac{dy_{1}^{-}}{2\pi}\, \frac{dy^{-}}{2\pi}\,
     \frac{dy_{2}^{-}}{2\pi}\,
     e^{ixp^{+}y_{1}^{-}}\, e^{ix_{k}p^{+}y^{-}}\,
     e^{-i(x_{k}-x_{k'})p^{+}y_{2}^{-}}  \nonumber \\
&\ & \times \frac{1}{2}\,
     \langle p_{A} |  F_{\alpha}^{\ +}(y_{2}^{-})\, \bar{\psi}_q(0)\,
           \gamma^+\, \psi_q(y_{1}^{-})\, F^{+\alpha}(y^{-})
     | p_{A}\rangle\ .
\label{e19}
\end{eqnarray}
In Eq.~(\ref{e19}), $F^{+\alpha}=F^{\beta\alpha}n_{\beta}$, and
$F^{\beta\alpha}$ is the field strength, and vector
$n_{\beta}=\delta_{\beta +}$.

In following sections, we show how to perform the integrations over
parton momentum fractions, and evaluate the partonic parts
$(\partial^2/\partial k_T^{\alpha}\partial k_T^{\beta})\, \bar{H}$
for different subprocesses.

\subsection{Leading pole approximation}
\label{subsec:3b}

The double scattering contribution defined in Eq.~(\ref{e18}) depends
on integrations over three partonic momentum fractions $x$, $x_k$,
$x_{k'}$.  If all partons in Fig.~\ref{fig4} carry some finite
momentum fractions, the oscillations of the exponentials in
the matrix element $T$ defined in Eq.~(\ref{e19}) will destroy any
nuclear size enhancement that could come from the $y$ integrations.
However, even at the lowest order, we find that there are some Feynman
diagrams which have two poles corresponding to zero momentum fraction
partons, and, these poles are not pinched.  Therefore, two of the
three parton momentum fractions can be integrated explicitly by
contour integration.  These integrations will eliminate two
exponentials, and thus, the corresponding $y$ integration could
provide the nuclear size enhancement up to $A^{2/3}$.  But, in terms
of double scattering picture, if we require two soft field operators
to come from the same nucleon, we will get the familiar $A^{1/3}$
enhancement.

Of course, there are double scattering diagrams without such poles,
but we expect $A^{\alpha}$ ($\alpha >1$) dependence only when the
poles are present.
In this paper, we evaluate only diagrams that have such poles, and we
call our results at leading pole approximation.

In order to perform the integration of momentum fractions, it is
convenient to rewrite the double scattering contribution defined in
Eq.~(\ref{e18}) as
\begin{eqnarray}
E_{l}\frac{d\sigma^{(D)}_{qA\rightarrow\gamma}}{d^{3}l}
&=& \frac{1}{2x's}\,
    \int \frac{dy_{1}^{-}}{2\pi}\, \frac{dy^{-}}{2\pi}\,
      \frac{dy_{2}^{-}}{2\pi}\,
\frac{1}{2}\,
     \langle p_{A} |  F_{\alpha}^{\ +}(y_{2}^{-})\, \bar{\psi}_q(0)\,
           \gamma^+\, \psi_q(y_{1}^{-})\, F^{+\alpha}(y^{-})
     | p_{A}\rangle \nonumber \\
&\times &
    \left(-\frac{1}{2}g^{\alpha\beta}\right)
\left[\, \frac{1}{2}\, \frac{\partial^{2}}
                   {\partial k_{T}^{\alpha} \partial k_{T}^{\beta}}\,
    H(y^-_1,y^-,y^-_2,k_T=0,p,l)\, \right]\ .
\label{e20}
\end{eqnarray}
In Eq.~(\ref{e20}), the modified partonic part $H$ is defined as
\begin{eqnarray}
H(y^-_1,y^-,y^-_2,k_T,p,l)
&=& \int dx\, dx_{k}\, dx_{k'}\,
     e^{ixp^{+}y_{1}^{-}}\, e^{ix_{k}p^{+}y^{-}}\,
     e^{-i(x_{k}-x_{k'})p^{+}y_{2}^{-}}  \nonumber \\
&\ & \times \bar{H}(x'p',x,x_k,x_{k'},k_T,p,l)\ ,
\label{e21}
\end{eqnarray}
where the partonic part $\bar{H}$ is given by diagrams shown in
Fig.~\ref{fig4}.  It is clear from Eq.~(\ref{e21}) that all integrals
of momentum fractions can now be done explicitly without knowing the
details of the multiparton matrix elements.

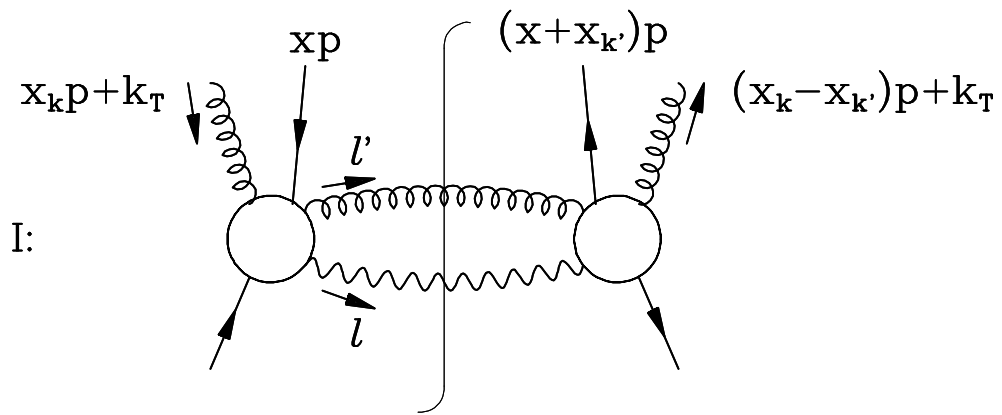
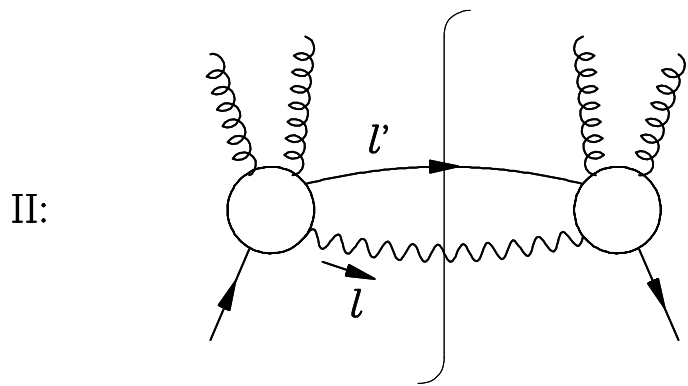
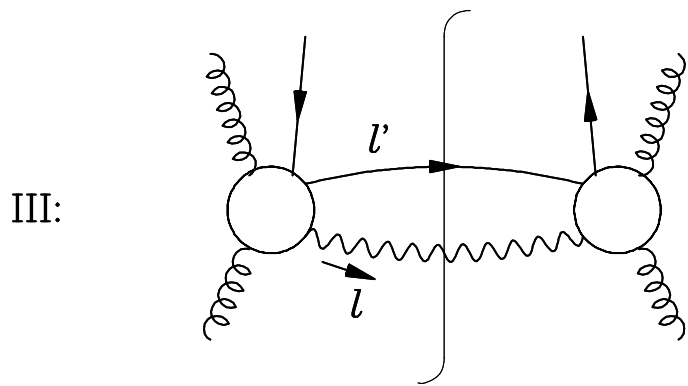
\begin{figure}
\caption{Three types of leading order Feynman diagrams contribute to
the double scattering.
a) Type-I: ``Annihilation'' diagrams corresponding to the
two-quark-two-gluon matrix element;
b) Type-II: ``Compton'' diagrams corresponding to the four-gluon matrix
element;
c) Type-III: ``Compton'' diagrams
corresponding to the two-quark-two-gluon matrix element.}
\label{fig3}
\end{figure}

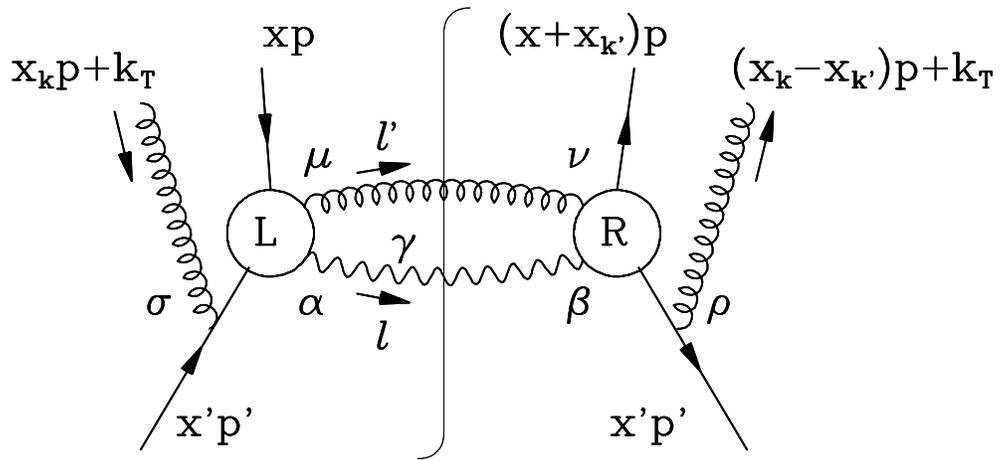
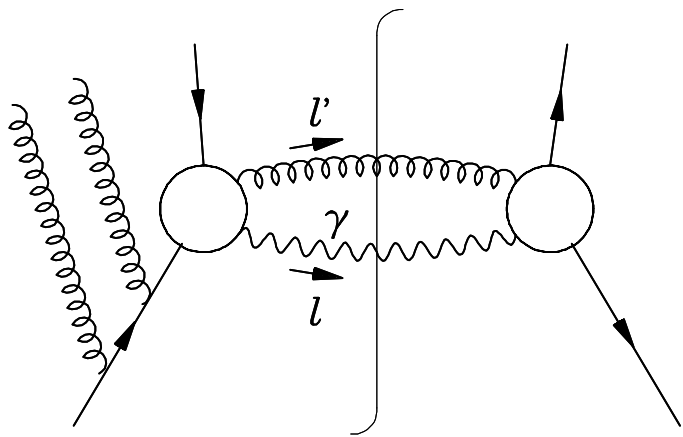
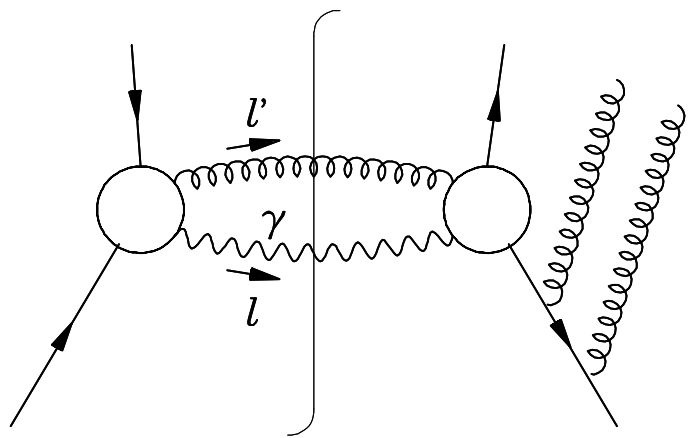
\begin{figure}
\caption{\sloppy
Feynman diagrams for the ``Annihilation'' diagrams
corresponding to the two-quark-two-gluon matrix element: a) real
diagrams, b) and c) interference diagrams.}
\label{fig4}
\end{figure}

Consider the diagram shown in Fig.~\ref{fig4}a.  The final state
photon-gluon two particle phase space can be written as
\begin{equation}
\Gamma=\frac{1}{8\pi^2}\, \frac{1}{x's+u}\,
\delta\left(x+x_k+\frac{x't}{x's+u}+\frac{-k_T^2-2k_T\cdot l}{x's+u}
      \right)\ .
\label{e22}
\end{equation}
In deriving Eq.~(\ref{e22}), we have omitted the factor $d^3l/E_l$,
due to the definition of the invariant cross section
(e.g., see Eq.~(\ref{e20})).
Using Eq.~(\ref{e22}), the contribution to $\bar{H}$ from the diagram
shown in Fig.~\ref{fig4}a can be expressed as
\begin{eqnarray}
\bar{H}_{I-a} &=&
\frac{\alpha_s}{2\pi}\, C_{I}\, \frac{1}{x's+u}\,
      \hat{H}_{I-a}(x,x_k,x_{k'}) \nonumber \\
&\times &
\frac{1}{x_k-x_{k'}-\frac{k_T^2}{x's}-i\epsilon}\,
\frac{1}{x_k-\frac{k_T^2}{x's}+i\epsilon}\,
\delta\left(x+x_k+\frac{x't}{x's+u}+\frac{-k_T^2-2k_T\cdot l}{x's+u}
      \right)\ ,
\label{e23}
\end{eqnarray}
where the subscript ``$I$-$a$'' has following convention: ``$I$''
stands for the type-$I$ subprocess, shown in Fig.~\ref{fig3}a; ``$a$''
for the real contribution, corresponding to diagrams in
Fig.~\ref{fig4}a.  In Eq.~(\ref{e23}), the factor $C_I$ is an overall
color factor for the type-$I$ subprocess.  The function
$\hat{H}_{I-a}$ in Eq.~(\ref{e23}) is
given by
\begin{equation}
\hat{H}_{I-a}=\frac{1}{4}\, \frac{1}{x's}\,
{\rm Tr}\left[\gamma\cdot(x'p'+k_T)\gamma\cdot p\, \gamma\cdot(x'p'+k_T)
         R_{I-a}^{\beta\nu}\gamma\cdot p\, L_{I-a}^{\alpha\mu}\right]
\left(-g_{\alpha\beta}\right)\left(-g_{\mu\nu}\right)\ ,
\label{e24}
\end{equation}
where $R_{I-a}^{\beta\nu}$ and $L_{I-a}^{\alpha\mu}$ are the right and
left blob, respectively, as shown in Fig.~\ref{fig4}a.  These blobs
include all possible tree Feynman diagrams with the external partons
shown in the figure.  Substituting Eq.~(\ref{e23}) into Eq.~(\ref{e21}),
we obtain
\begin{eqnarray}
H_{I-a}&=&
\frac{\alpha_s}{2\pi}\, C_{I}\, \frac{1}{x's+u}\,
\int dx_{k}\,e^{ix_{k}p^{+}(y^{-}-y^{-}_2)}\,
\frac{1}{x_k-\frac{k_T^2}{x's}+i\epsilon}\,
\int dx_{k'}\,e^{ix_{k'}p^{+}y^{-}_2}\,
\frac{1}{x_k-x_{k'}-\frac{k_T^2}{x's}-i\epsilon} \nonumber \\
&\ & \times
\int dx\, e^{ixp^{+}y^{-}_1}\,
\delta\left(x+x_k+\frac{x't}{x's+u}+\frac{-k_T^2-2k_T\cdot l}{x's+u}
      \right)\, \hat{H}_{I-a}(x,x_k,x_{k'}) \ .
\label{e25}
\end{eqnarray}
After performing $dx_k$ and $dx_{k'}$ by contour integration, and $dx$
by the $\delta$-function, we derive
\begin{eqnarray}
H_{I-a} &=&
(2\pi\alpha_s)\, C_{I}\, \frac{1}{x's+u}\,
e^{i\bar{x}p^{+}y^{-}_1}\,
e^{i\left(k_T^2/x's\right)p^{+}(y^{-}-y^{-}_2)} \nonumber \\
&\ & \times
\theta(-y^{-}_2)\, \theta(y^{-}_1-y^{-})\,
\hat{H}_{I-a}(\bar{x},x_k,x_{k'})\ ,
\label{e26}
\end{eqnarray}
where the $\theta$-functions result from the contour integrations,
and the momentum fractions for the function $\hat{H}_{I-a}$ are
defined as
\begin{mathletters}
\label{e27}
\begin{eqnarray}
\bar{x}&=&-\frac{1}{x's+u}\left[x't+\frac{u}{x's}k_T^2
-2k_T\cdot l\right]\ ;
\label{e27a} \\
x_k&=&\frac{k^2_T}{x's}\ ;
\label{e27b} \\
x_{k'}&=&0 \ ;
\label{e27c} \\
x&=&-\frac{x't}{x's+u}\ .
\label{e27d}
\end{eqnarray}
\end{mathletters}
\noindent Similarly, we derive contribution from the diagram shown in
Fig.~\ref{fig4}b as
\begin{eqnarray}
H_{I-b} &=&
(2\pi\alpha_s)\, C_{I}\, \frac{1}{x's+u}\,
e^{ixp^{+}y^{-}_1}\,
e^{i\left(k_T^2/x's\right)p^{+}(y^{-}-y^{-}_2)} \nonumber \\
&\ & \times
\theta(y^{-}_2-y^{-})\, \theta(y^{-}_1-y^{-}_2)\,
\hat{H}_{I-b}(x,x_k,x_{k'})\ ,
\label{e28}
\end{eqnarray}
where $x$, $x_k$ and $x_{k'}$ are also defined in
Eq.~(\ref{e27}).  Similarly to Eq.~(\ref{e24}), the partonic part
$\hat{H}_{I-b}$ is given by
\begin{equation}
\hat{H}_{I-b}=\frac{1}{4}\,
{\rm Tr}\left[\gamma\cdot(x'p')
         R_{I-b}^{\beta\nu}\gamma\cdot p\, L_{I-b}^{\alpha\mu}\right]
\left(-g_{\alpha\beta}\right)\left(-g_{\mu\nu}\right)\ .
\label{e29}
\end{equation}
The diagram shown in Fig.~\ref{fig4}c has following contribution
\begin{eqnarray}
H_{I-c} &=&
(2\pi\alpha_s)\, C_{I}\, \frac{1}{x's+u}\,
e^{ixp^{+}y^{-}_1}\,
e^{i\left(k_T^2/x's\right)p^{+}(y^{-}-y^{-}_2)} \nonumber \\
&\ & \times
\theta(y^{-}-y^{-}_2)\, \theta(-y^{-})\,
\hat{H}_{I-b}(x,x_k,x_{k'})\ .
\label{e30}
\end{eqnarray}
In deriving Eq.~(\ref{e30}), we used the fact that the partonic part
$\hat{H}_{I-c}=\hat{H}_{I-b}$ when $x_k$ and $x_{k'}$ are evaluated at
the same values as listed in Eq.~(\ref{e27}).

Combining $H_{I-a}$, $H_{I-b}$ and $H_{I-c}$ (given in
Eqs.~(\ref{e26}), (\ref{e28}), and (\ref{e30}), respectively)
together, we obtain the total contribution to $H$, defined in
Eq.~(\ref{e21}), from the type-$I$ diagrams shown in Fig.~\ref{fig3}a,
\begin{eqnarray}
H_{I} &=& H_{I-a} + H_{I-b} + H_{I-c} \nonumber \\
&=&
(2\pi\alpha_s)\, C_{I}\, \frac{1}{x's+u}\,
e^{i\left(k_T^2/x's\right)p^{+}(y^{-}-y^{-}_2)}\,
\theta(-y^{-}_2)\, \theta(y^{-}_1-y^{-})  \nonumber \\
&\ & \times
\left[ e^{i\bar{x}p^{+}y^{-}_1}\,
       \hat{H}_{I-a}(\bar{x},x_k,x_{k'})
     - e^{ixp^{+}y^{-}_1}\,
       \hat{H}_{I-a}(x,x_k,x_{k'}) \right] \ .
\label{e31}
\end{eqnarray}
All momentum fractions in Eq.~(\ref{e31}) are evaluated at the values
defined in Eq.~(\ref{e27}).
In deriving Eq.~(\ref{e31}), we have dropped a term proportional to
\[
[\,\theta(-y^{-}_2)\, \theta(y^{-}_1-y^{-})
-\theta(y^{-}_2-y^{-})\, \theta(y^{-}_1-y^{-}_2)
-\theta(y^{-}-y^{-}_2)\, \theta(-y^{-})\,]
\longrightarrow 0\ .
\]
This is because of the phase\ exp$[ixp^+y^-]$ which effectively
restricts $y^{-}_1\sim 1/(xp^+) \rightarrow 0$.  Physically, it means
that all $y$ integrations in such term are localized, and therefore,
will not give any large nuclear size enhancement.

By substituting Eq.~(\ref{e31}) into Eq.~(\ref{e20}), we can obtain
the lowest order double scattering contribution from the type-$I$
diagrams shown in Fig.~\ref{fig3}a.  One important step in getting the
final result is taking the derivative with respect to $k_T$ as defined
in Eq.~(\ref{e20}).  Comparing Eq.~(\ref{e31}) with Eq.~(\ref{e20}),
and observing that
\begin{equation}
\left[ e^{i\bar{x}p^{+}y^{-}_1}\,
       \hat{H}_{I-a}(\bar{x},x_k,x_{k'})
     - e^{ixp^{+}y^{-}_1}\,
       \hat{H}_{I-a}(x,x_k,x_{k'}) \right]_{k_T=0} = 0 \ ,
\label{e32}
\end{equation}
we found that the derivatives on the exponential\
exp$[i\left(k_T^2/x's\right)p^{+}(y^{-}-y^{-}_2)]$
do not contribute, and that we can therefore set
exp$[i\left(k_T^2/x's\right)p^{+}(y^{-}-y^{-}_2)]=1$ in
Eq.~(\ref{e31}).  Substituting Eq.~(\ref{e31}) into Eq.~(\ref{e20}),
and use Eq.~(\ref{e13a}), we obtain
\begin{eqnarray}
E_{l}\frac{d\sigma^{(D)}_{I}}{d^{3}l}
&=&
\alpha_{em}(4\pi\alpha_s)^2\,e_q^2\, C_I\,
\frac{1}{2x's}\, \frac{1}{x's+u}   \nonumber \\
&\times &
\left(-\frac{1}{2}g^{\alpha\beta}\right)\,
\frac{1}{2}\, \frac{\partial^{2}}
                   {\partial k_{T}^{\alpha} \partial k_{T}^{\beta}}
\left[\, T_q(\bar{x},A)\, \hat{H}_{I-a}(\bar{x},x_k,x_{k'})
        -T_q(x,A)\, \hat{H}_{I-b}(x,x_k,x_{k'})\, \right] \ ,
\label{e33}
\end{eqnarray}
where $\sigma^{(D)}_I$ stands for the double scattering contribution
from the type-$I$ subprocess shown in Fig.~\ref{fig3}a.  It is
important to note that although the interference diagrams shown in
Fig.~\ref{fig4}b and Fig.~\ref{fig4}c are important in driving
Eq.~(\ref{e33}), the final
result depends only on the real diagram shown in Fig.~\ref{fig4}a.
That is, the double scattering picture is preserved.  The role of
interference diagrams is to take care of the infrared sensitivities of
the short-distance hard parts.

\subsection{Final factorized form}
\label{subsec:3c}

The derivatives with respect to $k_T$ in Eq.~(\ref{e33}) are
straightforward.  It is most convenient to reexpress derivatives with
respect to $k_T$ in terms of derivatives with
respect to $\bar{x}$ or $x$.  After working out the derivatives, we
obtain
\cite{LQS}
\begin{eqnarray}
&&E_{l}\frac{d\sigma^{(D)}_{I}}{d^{3}l} =
\alpha_{em}(4\pi\alpha_s)^2\,e_q^2\,
\frac{1}{2x's}\, \frac{1}{x's+u}\, H_{q\bar{q}} \nonumber \\
&&\ \ \times 2\,
\Bigg\{\, \left[ \frac{\partial^{2}}{\partial x^{2}}
\left( \frac{T_{q}(x,A)}{x} \right) \right]
\left( \frac{l_{T}^{2}}{(x's+u)^{2}} \right)
+\left[ \frac{\partial}{\partial x}
\left( \frac{T_{q}(x,A)}{x} \right) \right]
\left( \frac{-u}{x's(x's+u)} \right)\, \Bigg\} \ ,
\label{e34}
\end{eqnarray}
where $x$ is given in Eq.~(\ref{e27d}), and where the partonic hard
part $H_{q\bar{q}}$ is defined as
\begin{equation}
H_{q\bar{q}} = C_I \, x\, \hat{H}_{I-a}(x,x_k=0,x_{k'}=0)\ .
\label{e35}
\end{equation}

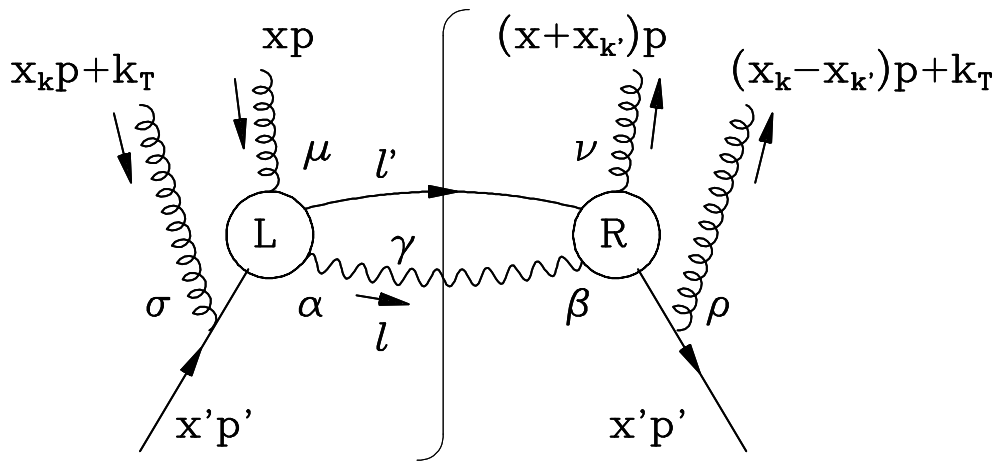
\begin{figure}
\caption{The real ``Compton'' diagrams corresponding to the
four-gluon matrix element.}
\label{fig5}
\end{figure}

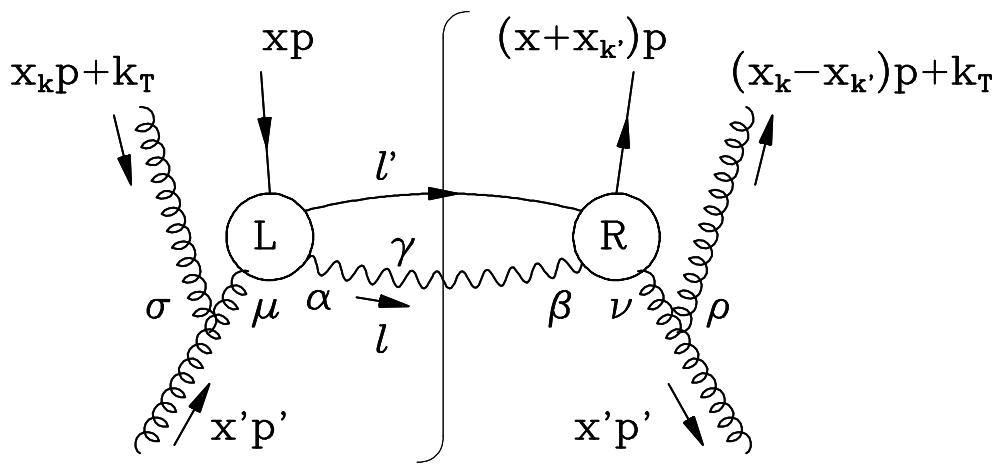
\begin{figure}
\caption{The real ``Compton'' diagrams
corresponding to the two-quark-two-gluon matrix element.}
\label{fig6}
\end{figure}

Following the same derivation, we obtain contributions from the
type-$II$ and type-$III$ diagrams shown in Fig.~\ref{fig3}.  For the
type-$II$ diagrams, as sketched in Fig.~\ref{fig3}b, we have
\begin{eqnarray}
&&E_{l}\frac{d\sigma^{(D)}_{II}}{d^{3}l} =
\alpha_{em}(4\pi\alpha_s)^2\,e_q^2\,
\frac{1}{2x's}\, \frac{1}{x's+u}\, H_{g} \nonumber \\
&&\ \ \times 2\,
\Bigg\{\, \left[ \frac{\partial^{2}}{\partial x^{2}}
\left( \frac{T_{g}(x,A)}{x} \right) \right]
\left( \frac{l_{T}^{2}}{(x's+u)^{2}} \right)
+\left[ \frac{\partial}{\partial x}
\left( \frac{T_{g}(x,A)}{x} \right) \right]
\left( \frac{-u}{x's(x's+u)} \right)\, \Bigg\} \ .
\label{e36}
\end{eqnarray}
In Eq.~(\ref{e36}), the partonic hard part $H_{g}$ is defined as
\begin{equation}
H_{g} = C_{II} \, \hat{H}_{II-a}(x,x_k=0,x_{k'}=0)\ ,
\label{e37}
\end{equation}
where $C_{II}$ is the overall color factor for the type-$II$ diagrams,
and $\hat{H}_{II-a}$ is given by the real diagrams shown in
Fig.~\ref{fig5}, and defined as
\begin{equation}
\hat{H}_{II-a}=\frac{1}{4}\,
{\rm Tr}\left[\gamma\cdot(x'p'+x_kp+k_T)
         R_{II-a}^{\beta\nu}\, \gamma\cdot l'\,
         L_{II-a}^{\alpha\mu}\right]
\left(-g_{\alpha\beta}\right)\left(-g_{\mu\nu}\right)\ ,
\label{e38}
\end{equation}
where $l'=x'p'+(x+x_k)p-l$ is the momentum carried by the quark going
to final state.  Similarly, for the type-$III$ diagrams, as sketched
in Fig.~\ref{fig3}c, we obtain
\begin{eqnarray}
&&E_{l}\frac{d\sigma^{(D)}_{III}}{d^{3}l} =
\alpha_{em}(4\pi\alpha_s)^2\,e_q^2\,
\frac{1}{2x's}\, \frac{1}{x's+u}\, H_{q} \nonumber \\
&&\ \ \times 2\,
\Bigg\{\, \left[ \frac{\partial^{2}}{\partial x^{2}}
\left( \frac{T_{q}(x,A)}{x} \right) \right]
\left( \frac{l_{T}^{2}}{(x's+u)^{2}} \right)
+\left[ \frac{\partial}{\partial x}
\left( \frac{T_{q}(x,A)}{x} \right) \right]
\left( \frac{-u}{x's(x's+u)} \right)\, \Bigg\} \ .
\label{e39}
\end{eqnarray}
The partonic hard part $H_{q}$ in Eq.~(\ref{e39}) is defined as
\begin{equation}
H_{q} = C_{III} \, x\, \hat{H}_{III-a}(x,x_k=0,x_{k'}=0)\ ,
\label{e40}
\end{equation}
where $C_{III}$ is the overall color factor for the type-3 diagrams,
and $\hat{H}_{III-a}$ is given by the real diagrams shown in
Fig.~\ref{fig6}, and defined as
\begin{equation}
\hat{H}_{III-a}=\frac{1}{4}\,
{\rm Tr}\left[\gamma\cdot p\,
         R_{III-a}^{\beta\nu}\, \gamma\cdot l'\,
         L_{III-a}^{\alpha\mu}\right]
\left(-g_{\alpha\beta}\right)\left(-g_{\mu\nu}\right)\ ,
\label{e41}
\end{equation}
where $l'$ is the same as that defined after Eq.~(\ref{e38}).

The partonic short-distance hard parts, defined in Eqs.~(\ref{e35}),
(\ref{e37}) and (\ref{e40}), can be easily evaluated by calculating
the corresponding Feynman diagrams.  Our results were presented in
Eq.~(\ref{e14}) in Sec.~\ref{sec:2b}.

After convoluting Eqs.~(\ref{e34}), (\ref{e36}), and (\ref{e39}) with
the corresponding parton distributions from the beam, we obtain the
complete analytical expressions for the double scattering contribution
in hadron-nucleus collisions, which was presented in Eq.~(\ref{e11})
of Sec.~\ref{sec:2b}.


\section{Numerical Results and Discussions}
\label{sec:4}

In this section, we present our numerical results for the Cronin
effect in direct photon production.  We numerically evaluate the
nuclear dependence parameter $\alpha(l)$ defined in Eq.~(\ref{e7}) by
using our analytical results presented in Eqs.~(\ref{e8}) and
(\ref{e11}), and we also compare our numerical results with recent data
from Fermilab experiment E706 \cite{Marek}.

The nuclear dependence parameter $\alpha(l)$ defined in Eq.~(\ref{e7})
depends on contributions from both single scattering and double
scattering.  All these contributions depend on the nonperturbative
parton distributions or multi-parton correlation functions.  In
deriving following numerical results, the Set 1 pion distributions of
Ref.~\cite{Pion} are used for pion beams; and the CTEQ3L parton
distributions of Ref.~\cite{CTEQ3} are used for free nucleons.  The
twist-4 multi-parton correlation functions defined in Eq.~(\ref{e13})
have not been well-measured yet.  By comparing the definition of
these twist-4 correlation functions with the normal twist-2 parton
distributions \cite{CS}, authors of Ref.~\cite{LQS} proposed
following approximate expressions for the twist-4 correlation
functions,
\begin{equation}
T_{i}(x,A)=\lambda^{2}\, A^{1/3}\, f_{i/A}(x,A)
\label{e42}
\end{equation}
where $i=q,\bar{q}$, and $g$.  The $f_{i/A}$ are the effective twist-2
parton distributions in nuclei, and the factor $A^{1/3}$ is
proportional to the size of nucleus.  The constant $\lambda^2$
has dimensions of [energy]$^2$ due to the
difference between twist-4 and twist-2 matrix elements.  The value of
$\lambda^2$ was estimated in Ref.~\cite{LQS2} by using the
\underline{measured} nuclear enhancement of the momentum imbalance of
two jets in photon-nucleus collisions \cite{E683,E609}, and was found
to be
\begin{equation}
\lambda^{2} \sim 0.05 - 0.1 \mbox{GeV}^{2}\ .
\label{e43}
\end{equation}
This value is not too far away from the naive expectation from the
dimensional analysis, $\lambda^2 \sim \Lambda_{\rm QCD}^2$.  In our
calculation below, we use $\lambda^2 = 0.1$~GeV$^2$.  Therefore, our
numerical results can be thought as the upper limit of the theoretical
predictions.

The $A^{1/3}$ dependence of the twist-4 multiparton correlation
functions, introduced in Eq.~(\ref{e42}), is not unique.  From the
definition of the correlation functions in Eq.~(\ref{e13}), the lack
of oscillation factors for both $y^-$ and $y^-_2$ integrals can in
principle give nuclear enhancement proportional to $A^{2/3}$.
The $A^{1/3}$ dependence is a result of the assumption that the
positions of two field strengths (at $y^-$ and $y^-_2$, respectively)
are confined within one nucleon.

In Eq.~(\ref{e42}), the effective nuclear parton distributions
$f_{i/A}$ should have the same operator definitions of the normal
parton distributions with free nucleon states replaced by the nuclear
states.  For a nucleus with $Z$ protons and atomic number $A$, we
define
\begin{equation}
f_{i/A}(x,A)=A \left( \frac{N}{A}f_{i/N}(x)
+\frac{Z}{A}f_{i/P}(x) \right)\, R_i^{\rm EMC}(x,A)\ ,
\label{e44}
\end{equation}
where $f_{i/N}(x)$ and $f_{i/P}(x)$ with $i=q,\bar{q},g$ are normal
parton distributions in a free neutron and proton, respectively; and
$N=A-Z$.  The factor $R_i^{\rm EMC}$ takes care of the EMC effect in
these effective nuclear parton distributions.  We adopted the $R^{\rm
EMC}$ from Ref.~\cite{BQV}, which fits the data well.  However, at
fixed target energies, the $x$ values covered by the direct photon
experiments are large and out of the nuclear shadowing region.  The
integration over $dx'$ in Eqs.~(\ref{e8}) and (\ref{e11}) averages out
the EMC effect from the large $x$ region.  Actually, one can
neglect the $R_i^{\rm EMC}$ in Eq.~(\ref{e44}).

Using the parton distributions and correlation functions introduced
above, and our analytic results presented in Eqs.~(\ref{e8}) and
(\ref{e11}), we can derive the nuclear dependence parameter
$\alpha(l)$, defined in Eq.~(\ref{e7}), \underline{without} any
further free parameter.

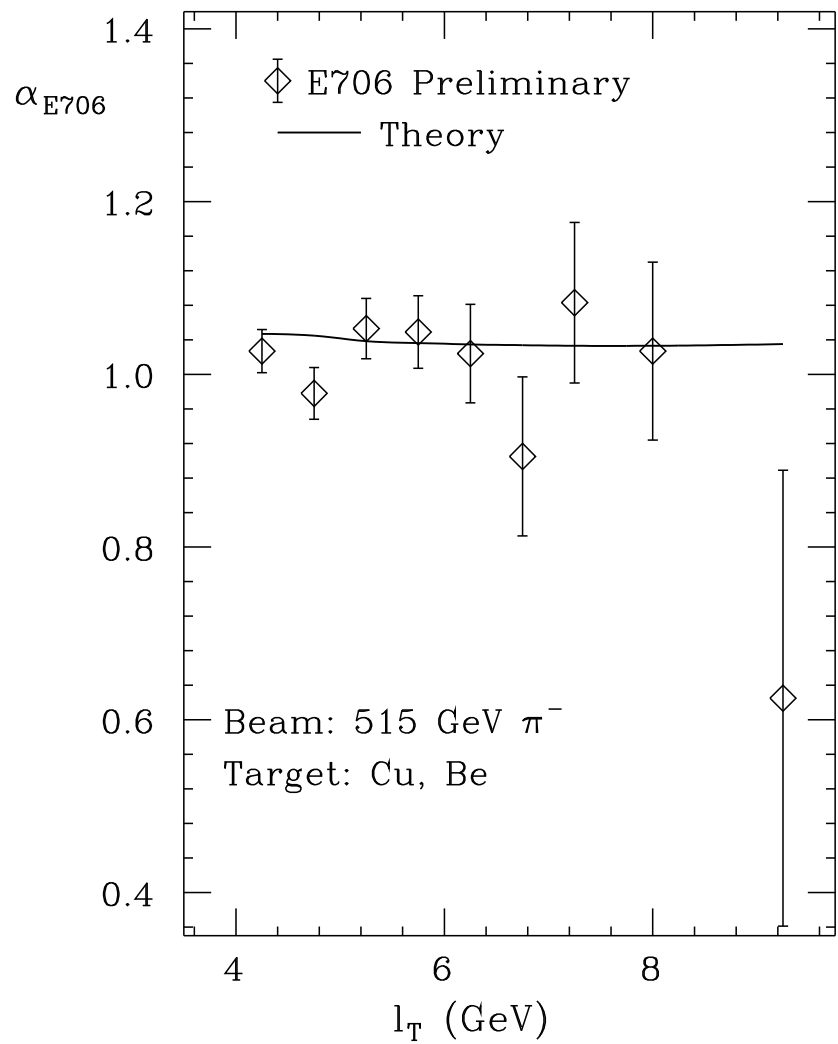
\begin{figure}
\caption{Cronin effect in direct photon production with 515 GeV
$\pi^-$ beam on Cu and Be targets.  The theory curve is from
Eq.~(\protect\ref{e45}), and the data is from Fermilab experiment E706
\protect\cite{Marek}.}
\label{fig7}
\end{figure}

In Fig.~\ref{fig7}, we compare our numerical predictions for the
nuclear dependence parameter with the recent experimental data from
Fermilab experiment E706 \cite{Marek}.  The $\alpha_{\rm E706}(l)$
presented in Fig.~\ref{fig7} is slightly different from that defined
in Eq.~(\ref{e7}).  E706 measured the direct photon cross sections
with the $\pi^-$ beam on two different targets: Cu($A=63.55$) and
Be($A=9.01$); and the $\alpha_{\rm E706}(l)$ was extracted according
to following definition
\begin{equation}
\frac{\sigma_{\rm Cu}(l)}{\sigma_{\rm Be}(l)} \equiv
\left(\frac{A_{\rm Cu}}{A_{\rm Be}} \right)^{\alpha_{\rm E706}(l)}
\ .
\label{e45}
\end{equation}
The beam energy is $p'=515$ GeV.  It is clear that the theoretical
calculation presented in this work is consistent with the data.

It is evident from Fig.~\ref{fig7} that the nuclear dependence parameter
$\alpha_{\rm E706}(l)$ is very close to the unity, or equivalently,
the Cronin effect for direct photon production is very small, and much
smaller than that observed in the single particle inclusive cross
sections \cite{Cronin}. One clear difference between the direct photon
and the single particle inclusive cross sections is that direct photon
production has only initial state multiple scattering, while the
single particle inclusive has both initial and final state multiple
scattering.  In addition, the single particle inclusive cross sections
depend on the parton-to-hadron fragmentation functions.

As pointed out in Ref.~\cite{LQS}, the multiple scattering
contribution is most important when the momentum fraction $x$ from the
nuclear correlation functions is large because of the derivatives with
respective to the $x$, which were introduced in Eq.~(\ref{e12}).
However, for direct photon production, the
kinematics do not fix all parton momentum fractions, and
leave one momentum fraction to be integrated, for example, $x'$
in Eqs.~(\ref{e8}) and (\ref{e11}).  Because of the steeply falling
nature of the distributions and correlation functions, the cross
sections in the central region ($x_F\sim 0$) for a given value of
$l_T$ is dominated by the distributions with momentum fractions
$x'\sim x\sim x_T =2l_T/\sqrt{s}$, which is less than 0.6 even for the
largest value of $l_T$ shown in Fig.~\ref{fig7}.  Therefore, the
double scattering contribution is relatively small because
the derivative terms are not significantly enhanced, and
consequently, $\alpha_{\rm E706}(l)$ is close to one.

In contrast, for inclusive single hadron production, the
parton-to-hadron fragmentation functions effectively shift the
contribution at a given $l_T$ to the large $x$ region because of all
fragmentation functions vanish when $z$ goes to 1.
Kinematically, direct photon production
corresponds to single particle production at $z=1$.  We, therefore,
expect that single hadron production has a larger Cronin effect than
the direct photon production at the exact same kinematics, even before
including contributions of final state multiple scattering.

In the case of a pion beam, the quark-antiquark ``annihilation''
subprocess, sketched in Fig.~\ref{fig1}a, dominates the production of
direct photons at the fixed target energy, due to the valence
antiquarks in the beam.  However, for a proton beam, the
quark-gluon ``Compton'' subprocess, as sketched in Fig.~\ref{fig1}b,
is more important for the production of direct photons.  Therefore,
direct photon production with a proton beam is more sensitive to the
proton gluon distributions.  In Figs.~\ref{fig8}, and
\ref{fig9}, we present our predictions of nuclear dependence parameter
$\alpha(l)$, defined in Eq.~(\ref{e7}), for a $\pi^-$ and proton beam,
respectively.

In Figs.~\ref{fig8}a and \ref{fig9}a, the nuclear dependence parameter
$\alpha(l)$, defined in Eq.~(\ref{e7}), is plotted as a function of
photon's transverse momentum $l_T$ at $x_F=0$.  In plotting these
figures, a 515 GeV beam energy and a Copper target was assumed, using
the same parton distributions and correlation functions as in
Fig.~\ref{fig7}.  As expected, and as found in
Fig.~\ref{fig7}, the value of $\alpha(l)$ is very close to
one for both pion and proton beams.  Changing a pion beam by a proton
beam does not affect the kinematics of the collisions.  The effective
values of parton momentum fractions from the nuclear target are the
same for both cases.  Therefore, as explained above, the values of
$\alpha(l)$ is close to one due to the fact that the effective
parton momentum fractions from the nuclear target are not large.

In order to enhance the contribution from the double scattering, we
need to look for events at large negative $x_F$, where the
effective values of parton momentum fractions from the target are
larger.  In Figs.~\ref{fig8}b and \ref{fig9}b, we plot the nuclear
dependence parameter $\alpha(l)$, defined in Eq.~(\ref{e7}), as a
function of $x_F$ at $l_T=6.0$ GeV.  The same beam, target and beam
energy were used.  It is clear that when $x_F$ becomes large and
negative, $\alpha(l)$ increases.  This is consistent with
the fact that the negatively larger values of $x_F$, the bigger
effective values of parton momentum fractions from the nuclear target,
and consequently, larger derivatives, defined in Eq.~(\ref{e12}).

Comparing Figs.~\ref{fig8}b and \ref{fig9}b, one finds that as $x_F$
decreases, the values of $\alpha(l)$ with a proton beam
increases much fast than that with a pion beam.  This is because the
Compton subprocess dominates the production of direct photons in
the case of a proton beam, and
the gluon distribution in a proton falls much faster than the
valence quark distributions as the momentum fraction increases.
The more rapidly falling gluon distribution produces larger derivative
terms, and therefore, larger values of $\alpha(l)$.  Future data from
Fermilab experiment E706 with a proton beam can test this feature.

In summary, using generalized factorization in QCD perturbation
theory, and using the method, developed in Ref.~\cite{LQS},
for calculating nuclear enhancements,
we demonstrated in this paper that the observed small Cronin
effect in direct photon production is consistent with the much larger
Cronin effect observed in single jet and single particle inclusive
cross sections.  We hope that the same method can be used to explain
the puzzle for the nuclear dependence in the momentum imbalance.  Data
on dijet momentum imbalance in photoproduction \cite{E683} and
hadroproduction \cite{E609} have shown strong nuclear dependence,
while much smaller nuclear dependence has been seen in momentum
imbalance of Drell-Yan pairs \cite{DrellYan}.


\section*{Acknowledgment}

We would like to thank George Sterman for helpful discussions.
The work was supported in part by the U.S. Department of Energy under
Grant Nos. DE-FG02-87ER40731 and DE-FG02-92ER40730.




\begin{figure}
\caption{Theoretical predictions of $\alpha(l)$ of
Eq.~(\protect\ref{e7}): a) as a function of photon's transverse
momentum at $x_F=0.0$; b) as a function of $x_F$ at $l_T=6.0$~GeV.
515 GeV $\pi^-$ beam and Copper target were used. }
\label{fig8}
\end{figure}

\begin{figure}
\caption{Theoretical predictions of $\alpha(l)$ of
Eq.~(\protect\ref{e7}), as in Fig.~{\protect\ref{fig8}}, but with a
proton beam. }
\label{fig9}
\end{figure}


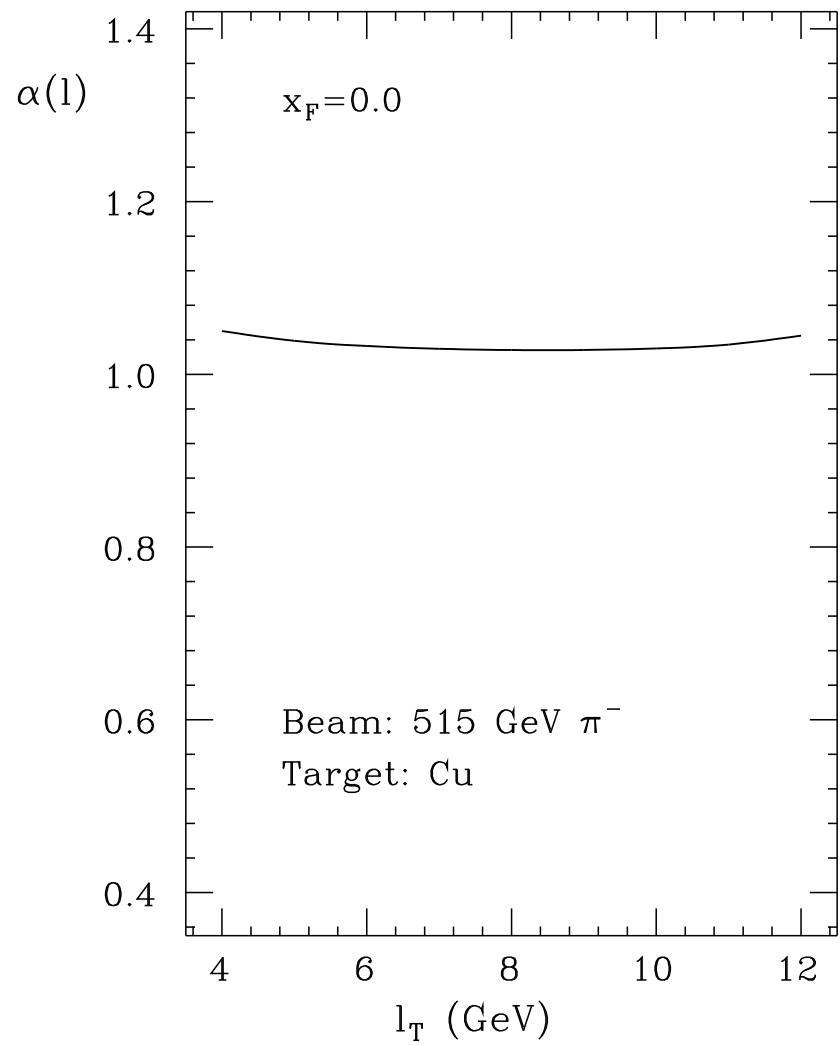
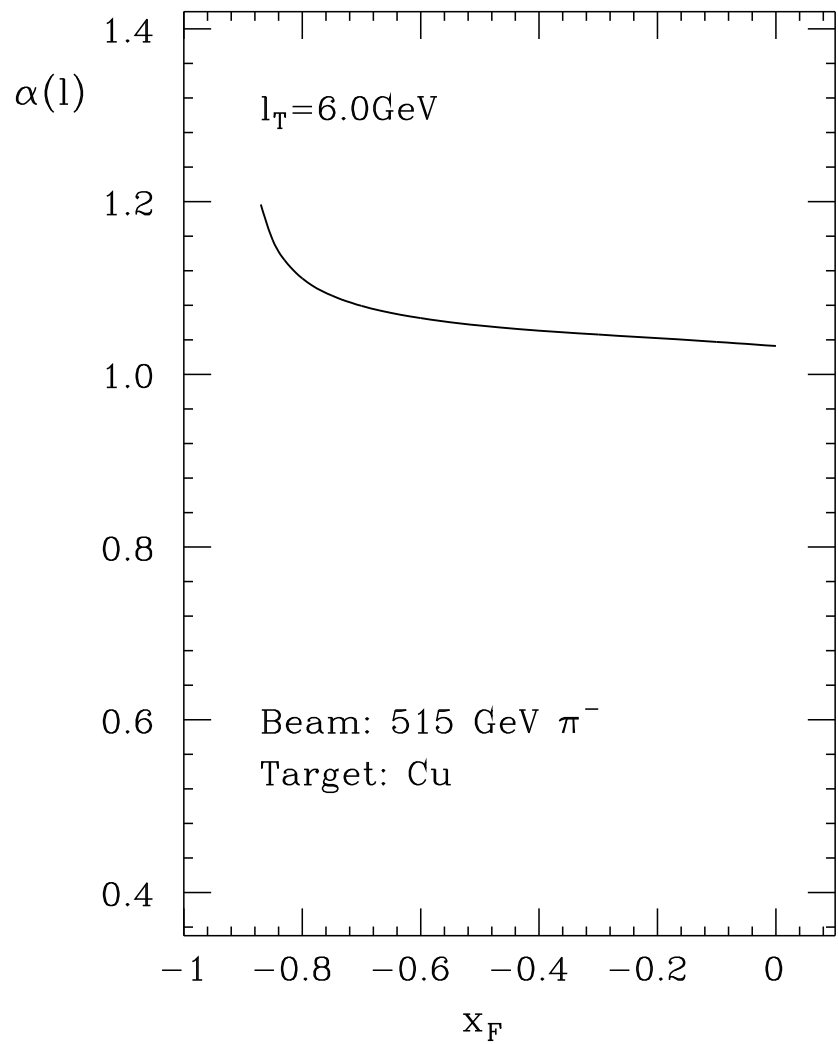
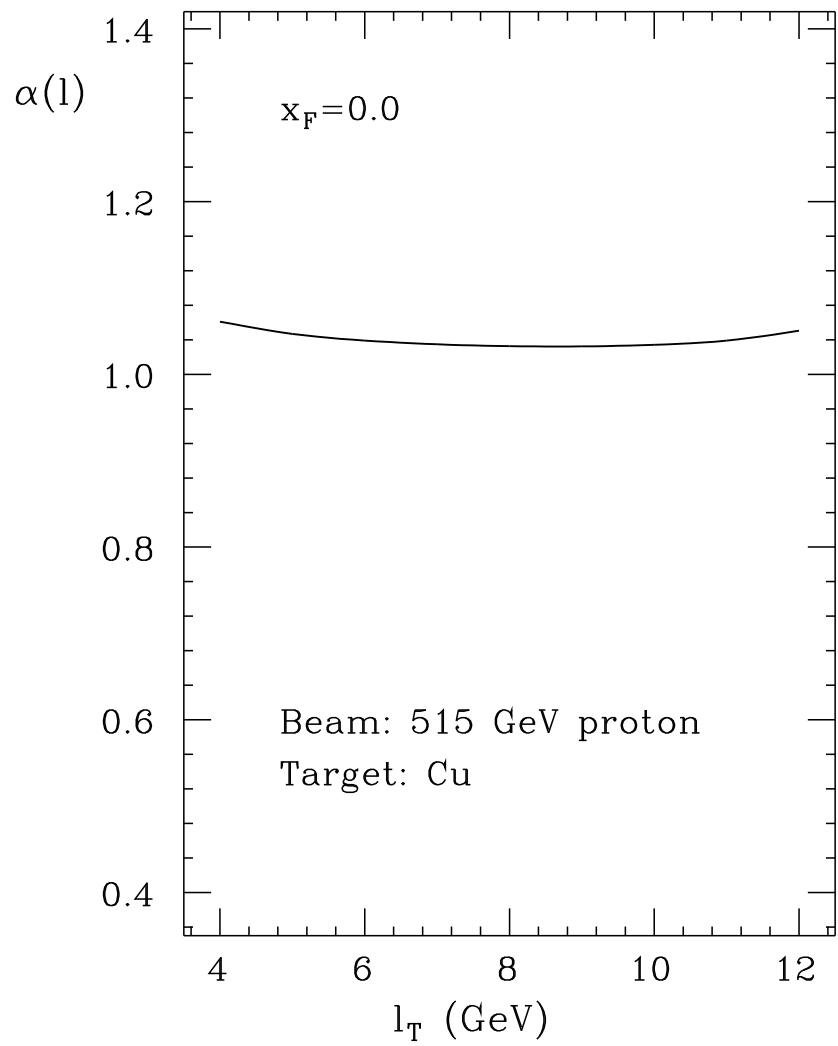
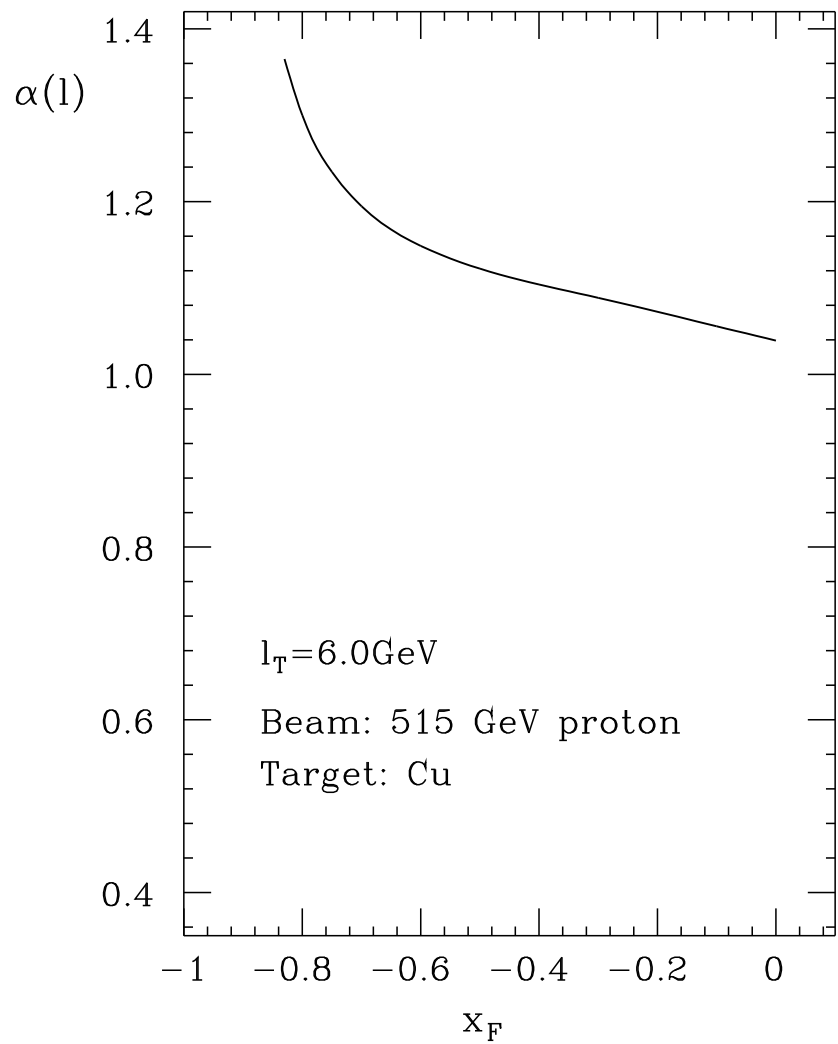
\end{document}